# Majorana bound states in d-wave superconductor-based hybrid systems

C. C. Tsuei

IBM Thomas J. Watson Research Center
Yorktown Heights, NY 10598, U.S.A.

## Abstract

For the search of Majorana fermions, we propose to employ a ramp-type Josephson edge junction between a normal metal such as Au, Ag or alloys of heavy metals and a topologically-trivial $d_{x^2-y^2}$ –wave high-$T_c$ superconductor $YBa_2Cu_3O_7$. The success in forming the elusive zero-energy Majorana bound states (MBS) at the junction relies crucially upon the strength of the Rashba spin-orbit coupling (RSOC) at the metal/superconductor interface, the tunneling direction, and the interface quality which includes a near zero tunnel barrier height. When these junction conditions are collectively satisfied, definitive signatures of Majorana fermions will emerge. After inspecting a vast amount of published data of junctions that are close to what we have proposed, we were able to find**:** 1) angular resonant Andreev reflection, and 2) quantized zero bias normalized differential conductance peak (ZBCP), as possible evidence for MBS formed at the junction interface. Encouraged by these findings, we have designed a MBS-supporting d-wave superconductor based π-loop heterostructure as a qubit platform. Such π-loops are characterized by a doubly degenerate half flux quantum vortex (HQV) and are topologically protected by a parity effect arising from an odd number of sign changes of the supercurrent circulating in the loop. The topological protection is further enhanced by the formation of RSOC induced Majorana bound states at the Josephson junction interface. The proposed MBS-plus-HQV qubit platform will be useful for demonstrating non-Abelian exchange statistics and for building a prototype topological quantum computer.

Ever since Ettore Majorana predicted in 1937 a spin half, charge neutral and real particle that is also its own anti-particle, the pursuit of such fundamental particles has never stopped and so far without success [1]. Recent advances in understanding the concepts of topological quantum number, topological order, non-Abelian brading states, and the discovery of topological insulators and topological superconductors [2 - 7] have shifted the search for the elusive Majoana particles to the field of condensed matter physics. Strictly speaking, the Majorana fermions in a condensed matter system, if ever can be found, are not fundamental particles like neutrinos, rather they have to be in the form of composite quasiparticles, called Majorana bound states (MBS) which possess the essential properties of the elementary particles predicted by Majorana. Interesting to mention is that the MBS can be encoded with information to enable non-Abelian exchange statistics for error-tolerant quantum computation [5, 8] a feat that could not be envisioned when the concept of Majorana particles was conceived 75 years ago.

A superconductor with its charge conjugated Bogoliubov quasiparticle (qp) excitations represents an ideal platform for the quest of MBS. However a topological system with non-Abelian characteristics needs not to be a superconductor. As shown by the work of Read and Green [9], the fractional Quantum Hall Effect (FQHE) material with filling level of 5/2 can be a candidate for topological non-trivial system. The topological nature of such a system can be derived equivalently from the qp excitation states of a superconductor with spinless $p_x+ip_y$ pairing symmetry [10]. Furthermore, its half flux quantum vortex (HQV, ½ $\Phi_0$ = h/4e) characteristic of a genuine p-wave superconductor is capable of hosting the MBS in the FQHE system. We will get back to this point later.

To construct a Majorana bound state in superconductors, one can first create a coherent superposition [11] of electron-like and hole-like qp excitations with energy E and spin index σ, with amplitudes $u_{E,\sigma}(\mathbf{r})$ and $v_{E,-\sigma}(\mathbf{r})$ respectively:

$$\gamma_{E,\sigma}^{\dagger} = \int d\mathbf{r}[u_{E,\sigma}(\mathbf{r})\Psi_{\sigma}^{\dagger}(\mathbf{r}) + v_{E,-\sigma}(\mathbf{r})\Psi_{-\sigma}(\mathbf{r})] \qquad (1)$$

provided that the most defining condition of the Majorana fermion state: the creation operator of such state is equal to its own annihilation counterpart, $\gamma_{E,\sigma}^{\dagger} = \gamma_{E,-\sigma}$, is satisfied. It is clear from Eq. (1) that the Majorana condition cannot be met trivially without imposing further constraints on the Bogoliubov quasiparticle states [3-7, 9, 10]. Since the spin space is independent of the spatial part of a quantum state in Hilbert space, it is probably easiest to accomplish this by lifting the spin degeneracy, meaning that only spin states with single species should be employed in the MBS formation. Based on Eq. (1), the Majorana requirements $u^*_{E,\sigma}(\mathbf{r}) = v_{E,-\sigma}(\mathbf{r})$ become $u^*_E(\mathbf{r}) = v_E(\mathbf{r})$, which lead to $\gamma_{-E}^{\dagger} = \gamma_E$ and $\gamma_0^{\dagger} = \gamma_0$ for zero-energy MBS. Guided with such constrains on qp excitations, numerous proposals have been suggested for the search of MBS in condensed matter systems [6 - 10]. A common theme threading through nearly all such search schemes is to employ spin orbit coupling (SOC) and / or Zeeman magnetic field to judiciously formulate the qp excitations for building the MBS in various superconducting systems[12–15]. With the SOC and /or Zeeman field included in the Boguliubov de Genne (BdG) Hamiltonian[11], the qp components $u$ and $v$ can be obtained by solving the BdG equations, subject to bandstructure, appropriate boundary conditions and constraints imposed by relevant symmetries (particle-hole, time reversal and pairing), for studying the existence and stability of MBS in various solid state systems.

Building on the earlier work in the field by Fu and Kane [12], Potter and Lee [13], we propose here to engineer d-wave superconductor based Josephson junction devices with a heavy metal (e.g. Au, Ag, Pb, Bi, and their alloys which are known for its strong SOC) as a part of the tunnel junction structure for searching the MBS at the junction interface. We will present some published experimental results of quantized zero-bias-conductance peak of one conductance quantum ($2e^2/h$) **and** that of resonant Andreev reflection as possible evidence for MBS in the proposed hybrid systems. Furthermore, we will proposed to use such heterostructures to build a Majorana qubit platform for demonstrating non-Abelian exchange statistics, constructing quantum logic gates for topological quantum computation.

To enlist the help of strong SOC in making the MBS, one could use a superconductor with an intrinsically large SOC effect deposited on a topological insulator [12]. Alternatively one could make a ferromagnetic material/topological superconductor junction exposed in a strong Zeeman field [6, 7]. Another alternative is to place a strongly spin-orbit coupled nanowire on a topologically trivial s-wave superconductor [14, 15]. Some interesting experimental results of such efforts have been published recently [16].

In this work, we use a nanometer thick layer of a heavy metal as a crucial component incorporated in a d-wave superconductor based Josephson junction for trapping the MBS at its junction interface. Then employing such SOC enhanced Josephson junctions to form a superconducting π-loop for hosting the Majorana fermions located at the junction interface. The idea of using Au to engineer a one-dimensional MBS device was studied by Potter and Lee [13], and recently by Yuan, Wong and Law [17]. A schematic of our hybrid junction structures is shown in Fig. 1(a) and (b).

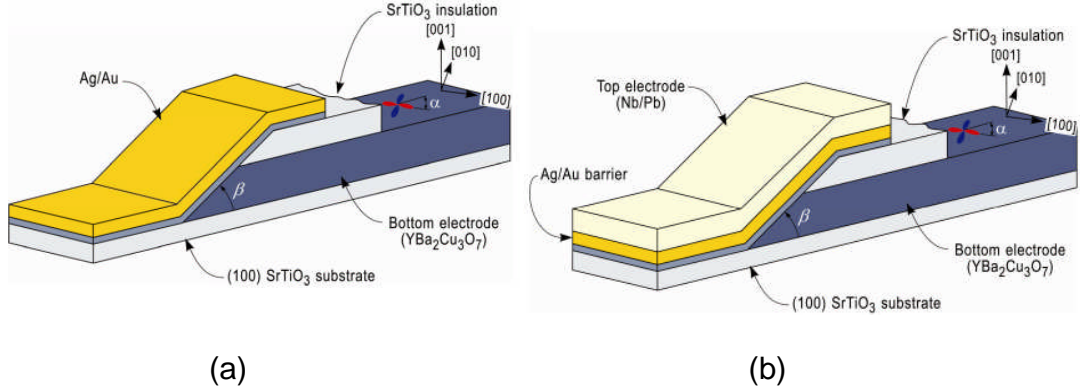

(a)                      (b)

---

Figure 1 A schematic of d-wave superconductor based heterostructures for the search of Majorana bound states: (a) a N/$S_d$ ramp-type Josephson edge junction, and (b) S/N/$S_d$ type ramp edge junction. The ramp type Josephson junctions involving d-wave superconductors ($S_d$) have been used for fundamental study of high-temperature superconductivity and its applications [18 – 20, 25]. As depicted in Fig. 1, the base electrode is a c-axis oriented d-wave superconductor $S_d$, e.g. YBa$_2$Cu$_3$O$_7$(YBCO), with an angle α between the principal in-plane crystallographic axis and the norm of the interface boundary line. The ramp-type interface makes an angle β with respect to the junction substrate. The ramp angle β is typically about 20 to 45 degrees. Due to the relatively large anisotropy in resistivity (the out-of-plane to in-plane resistivity is of the order of 50 or higher depending on the specific cuprate system and the sample quality), the planar tunneling is dominant over the c-axis tunneling. Then a ramp junction with α = 0 or π/4 corresponds to tunneling into [100] or [110] respectively. The heavy metal layer of Au or Ag can provide a strong Rashba SOC [21] at the junction interface. An ultrathin YBCO layer (5 nm thick) deposited and heat treated in-situ on the base electrode [19]. The function of this YBCO interlayer is to repair the rough surface and edge of the base electrode, and to enable high transparency at the N/ $S_d$ interface. A bright field electron transmission image of the Au/YBCO interface near the ramp edge area suggests a sub-nano-meter scale interface [19]. Such an atomically sharp interface between Au and YBCO was responsible for the observed high supercurrent density (>$10^4$ A/cm$^2$ at 4.2K) across the ramp junction. As we will see later, this is of utmost importance to forming the MBS in this kind of heterostructures.

---

The importance of SOC in the studies of topological systems was recognized from the beginning [3–7, 12-15]. In the surfaces and interfaces, SOC manifests itself as the Rashba effect [22, 21] which leads to a two-dimensional momentum dependent spin splitting of energy bands [22] and a mixing of orbital pairing symmetries as predicted by Gorkov and Rashba [21]. To describe the Rashba SOC at an interface, the basic Hamiltonian of a heterostructure system needs to include a term $H_R$ to reflect the existence of a structure inversion asymmetric potential: $H_R = \alpha_R (\mathbf{n} \times \mathbf{k}) \cdot \sigma$, where k is the electron wavevector along the in-plane direction of the junction boundary, **n** the unit vector perpendicular to the junction interface, and σ the Pauli spin matrices. The parameter $\alpha_R$ is the strength of the Rashba SOC. Recent experiments have found that it is interface (surface) specific and not totally dependent on its atomic number. Giant spin splitting can be achieved with a carefully engineered atomic layer structure in which the specific local electronic and structural configurations combined to produce an enhanced electric field gradient normal to the junction interface [22]. For example, the $\alpha_R$ values of the Au(111), Ag(111) and Bi(111) surfaces are 0.33[see ref. 22], 0.055[ Cercellier et al. (2006) see ref. 22 ] and 0.56 eV Å [ref 8 in Ast PRL, see ref 22] respectively. However, it was found that a giant Rashba spin splitting effect with $\alpha_R$ = 3.05 eV Å can be achieved at the surface of Bi/Ag(111) [C. Ast et al. (2007) in ref. 22]. Apparently there is no $\alpha_R$ data for Ag or Au on oxides available at this moment. Such information would be most relevant to our discussion later. However, based on the $\alpha_R$ information on Au(111) and Ag(111) surfaces, one can estimate that the Rashba spin splitting energy at Fermi level is of the order of 200 meV for Au/YBCO, 30 meV for Ag/YBCO at the interface layer.

With the spin degeneracy lifted by the Rashba SOC, the surface states at the junction can be described with two separate spin-polarized sub-bands near the Fermi level, $E_1(k)$ and $E_2(k)$ corresponding spin up↑ and spin down↓ respectively. Thus a MBS can be built by using the quasiparticle creation operator (in k-presentation) as follows:
$\gamma^\dagger(\mathbf{k}) = u_1(\mathbf{k})c_\uparrow^\dagger(\mathbf{k}) + u_2(\mathbf{k})c_\downarrow^\dagger(\mathbf{k}) + v_1(\mathbf{k})c_\uparrow(-\mathbf{k}) + v_2(\mathbf{k})c_\downarrow(-\mathbf{k})$, with $u_1(\mathbf{k}) = v_1^*(-\mathbf{k})$ and $u_2(\mathbf{k}) = v_2^*(-\mathbf{k})$. Then the Majorana condition, $\gamma^\dagger(\mathbf{k}) = \gamma(-\mathbf{k})$ is satisfied [23, 24]. With the spin degeneracy lifted by Rashba SOC, Majorana fremions can be in principle readily generated at the edge between $k_1$ and $k_2$ [23], and bound to a vortex [9, 10]. In practice, it is quite challenging to select an appropriate topological system with optimized experimental device structure and parameters to observe a definitive signature of the MBS predicted by theory. We will concentrate on the ramp-type $N/S_d$ Josephson junction based heterostructures, as depicted in Fig. 1, for our search for Majoran fermions. During the last twenty six years of the high-temperature superconductivity research, ramp junctions have been used for various fundamental studies (pairing symmetry, macroscopic quantum coherence phenomena, qp tunneling etc.)[25], SQUIDs and other superconducting device applications [20]. We will look first for possible evidence for MBS from the qp tunneling measurements already published in the literature.

Among all the MBS signatures, a zero-bias-conductance peak (ZBCP) of the differential conductance of a tunnel junction is probably the most straightforward to measure and understand. A vast amount of Andreev tunneling spectroscopy data [26], in agreement with theories [26,27], has established convincingly mid-gap zero-energy bound states formed on the surface (interface) of junctions containing a d-wave superconductor [26]. The formation of such Andreev bound states (ABS) originates from the constructive interference of the time-reversed incident and reflected quasiparticles experiencing a sign change of the order parameter fields (d-wave gap potentials). The existence of ABS manifests itself through a zero-bias conductance peak (ZBCP) in the in-plane (the $CuO_2$ plane) quasiparticle tunneling from the (110) but not (100) oriented surface, as predicted by the theory [26]. Although a ZBCP in the tunneling spectrum may be considered as signal of the existence of MBS at the junction interface. It may have origins arising from impurities and other sources that are irrelevant to MFs [26, 27]. A ZBCP was also sometimes observed, against the prediction of theory, in planar tunneling along the [100] and [103] directions due to microscopic-scale surface roughness, for example faceting occurred frequently on the surface (interface) of cuprate superconductors [28]. A typical surface or interface of cuprate superconductors is generally characterized by a random distribution of (100) and (110) facets [20, 28].

Of course, the existence of a zero energy mode of the MBS also presents itself as a ZBCP. There are theoretical studies predict that the resonant Andreev reflection through MBS in qp tunneling in a N/S junction should show a conductance change of $2e^2/h$ superposed on the conductance spectrum originated from standard qp tunneling [29, 30, 42, 6]. In our case of $N/S_d$ junctions, the standard conductance spectrum includes tunneling into the [110] and/or [100] directions and scattering from background sources.

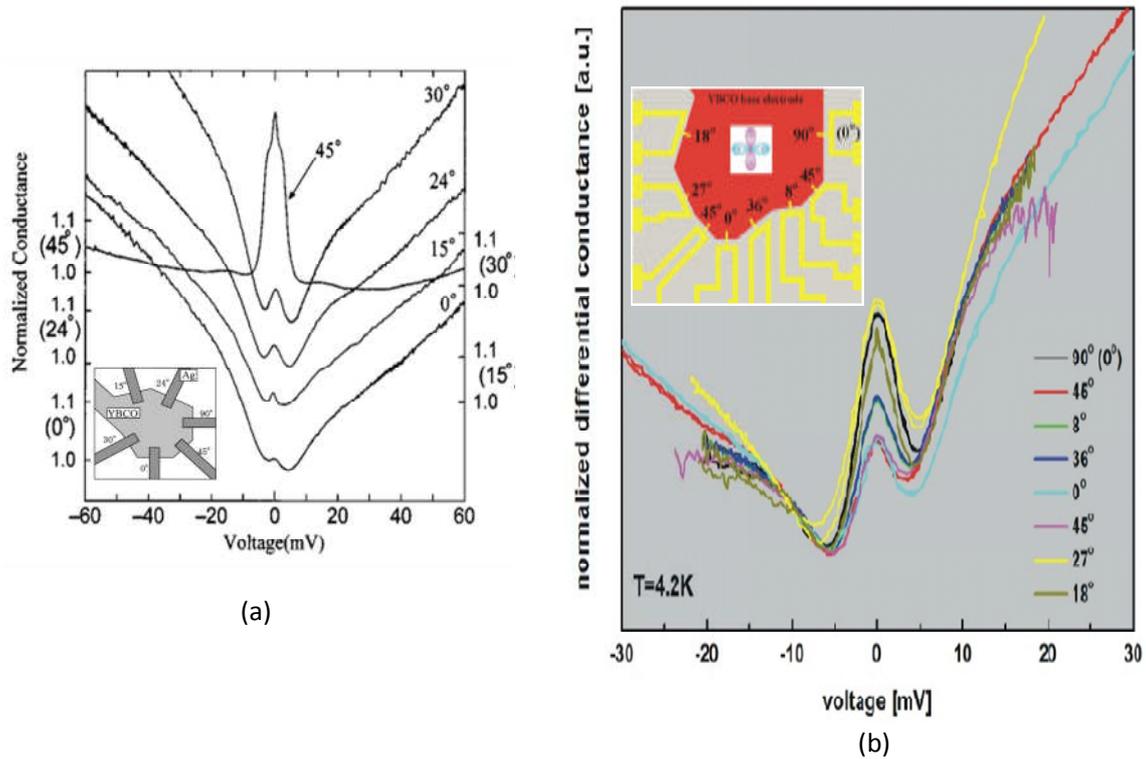

(a)

(b)

---
Figure 2 Angle-resolved normalized differential conductance (dI/dV) spectra of ramp junctions (a) Ag/YBCO [taken from Iguchi et al. ref.18] and (b) Au/YBCO [taken from the PhD thesis by C. Schuppler (Birmingham University 2009) [31]. The interface angles here are the same as angle α defined in Fig 1.

---

Shown in Fig. 2 are the results of two angle-resolved studies of qp tunneling across ramp junctions (a) Ag/YBCO taken from Iguchi et al.[18] and (b) Au/YBCO[taken from the PhD thesis by C. Schuppler (Birmingham University 2010)[31]. The angles marked in Fig. 2 are the same as the angle α defined in Fig. 1. The data in both figures reveal a small sharp peak superposed on top of a broad peak at zero bias. The fact that in Fig. 2 (a), this sharp ZBCP is pronounced only for tunneling into α = 45 deg demonstrates its character of resonant Andreev reflection. Furthermore, the small ZBCP comes with a bias voltage independent background conductance. All other samples in Fig. 2(a) are featured with an asymmetrical V-shaped spectrum background which can be attributed at least partially to a significant diffuse scattering in the tunneling across the junction [27]. The results in Fig. 2(b) also show a small sharp peak at zero bias voltage although occurred at a "wrong" angle (α = 18 degrees instead of 45 deg as shown in Fig. 2 (a)). In addition, the asymmetrical V-shaped conductance background is quite dominant in all the samples of this work. The unexpected resonant angle (α =18 deg) can be reconciled with the assumption that the junction interface is composed of both the (100) and (110) facets resulting an average value of <α> = π/8, very close to the observed value (α =18 degrees). If this explanation is correct, it means that in realistic macroscopic size samples, the overall resonant Andreev reflection associated with the MFs can be greatly suppressed due to a rough junction interface. The persistence of the asymmetrical V-shaped background conductance in Fig. 2(b) seems to support this conjecture.

The results in Fig. 2 above suggest possible evidence for MF induced resonant Andreev reflection across the angle-resolved ramp-type Ag/YBCO and Au/YBCO junctions. The magnitude of the conductance change at zero bias voltage, however, is far from being equal to the conductance quantum, $G_{MBS} = 2e^2/h$, expected by theory. Presumably this is due to the disorder effects at the junction interface.

With this in mind, let's inspect more data presented in Figure 3.

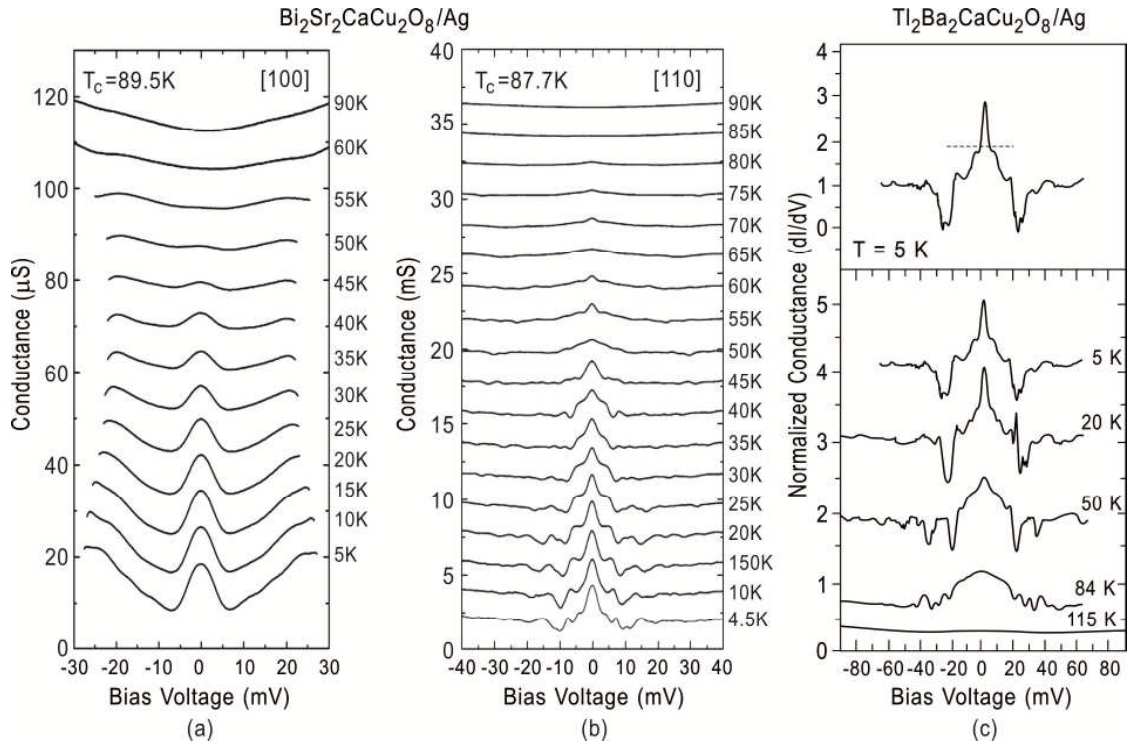

----------------------------------------------------------------------------------------------------
Figure 3 Quasiparticle tunneling spectra of $Bi_2Sr_2CaCu_2O_8$(single crystal)/Ag planar junctions with tunneling directions of [100] in (a), and [110] in (b) as a function of temperature. The data were taken from ref. 32 by I. Shigeta et al. (2002) [32]. The temperature dependent conductance spectra of a $Tl_2Ba_2CaCu_2O_8$ (c-axis oriented epitaxial film)/Ag planar junction, taken from F. Giubileo et al. (2002) [33], are shown in (c).
----------------------------------------------------------------------------------------------------

We will analyze these data with the following two considerations:

**1) Quality of the junction interface:**

Over the last two decades, tunneling spectroscopy has been developed to be a powerful technique for understanding the physics of electron transport across various junction interfaces, especially those involving high-temperature superconductors as a junction electrode [34, 35, 26 ]. During the course of these studies, it has become plentifully clear that a high-quality junction interface is a prerequisite for reliable data. The cuprate superconductors are characterized by short coherence length $\xi$ (e.g. in YBCO the in-plane

$\xi_{ab}$ ~ 2 nm, c-axis $\xi_c$ << 1 nm). In the Andreev reflection process, constructive interference occurs within the range of coherence length. This makes the point contact or planar tunneling spectroscopy an ultra sensitive surface/interface probe which collects intrinsic information as well as data arising from defects, impurities, disorder effects at the junction interface [36]. Furthermore, as pointed out by P. A. Lee, any MBS scheme using $p_x + ip_y$ pair state is particularly vulnerable to pair breaking by impurity and other defect scatterings [37]. Therefore to make sure the signal of the MBS that we are looking for is not masked and/or suppressed by such extrinsic effects, the experiment has to be done with samples of very high-quality junction interface. Notice that the tunnel junction electrodes used for the data shown in Fig. 2(a), and Fig. 3 all have a smooth or specially treated surface or junction interface. [18, 32 and 33].

**2) Generalized BTK model in Z = 0 limit:**

It is well-established theoretically and experimentally that the qp tunneling conductance spectrum of $N/S_d$ type Josephson junctions can be understood in terms of a generalized model [R. C. Hu (1994), S. Kashiwaya et al. (1995), in ref. 26] based on the theory of Blonder-Tinkham-Klapwijk (BTK) [38]. The original BTK model is for describing charge transport across the tunnel barrier of a $N/S_s$ junction where $S_s$ is a conventional s-wave superconductor. It treats the qp transport across the junction by using only one **dimensionless** and **temperature independent** parameter Z which is a measure of barrier-height. When the effects of Fermi level mismatching between the base and counter electrodes, lifetime broadening, - - - are taken into account, BTK and its generalized versions can still capture the essence of the phenomenology of electron transport in N/S and N/I/S tunnel junctions with only **one** effective barrier-height parameter Z. Such a remarkable success of the BTK model and its variants can be attributed to an effect of compensation among various corrections of the order $\Delta/E_F$ to the original BTK model [see ref. 39 for details]. Even for large $\Delta/E_F$ (likely the case of the cuprate superconductors), the overall correction is relatively small, and the conductance spectrum of the junction does not differ significantly from the original BTK model [39]. Shown in Fig. 4 (a) and (b) are the normalized differential conductance spectra of $N/S_d$ junctions based on a BTK model generalized for $d_{x2-y2}$-wave superconductors by Kashiyawa and Tanaka in 1995 [26]. The numerical results of this model are plotted for various barrier heights Z as a function of bias voltage in eV/$\Delta$, where $\Delta$ the superconducting energy gap, for tunneling in the direction of [110] as shown in Fig 4 (a), and in the direction of [100] displayed Fig. 4(b).

Fig. 4 (a)

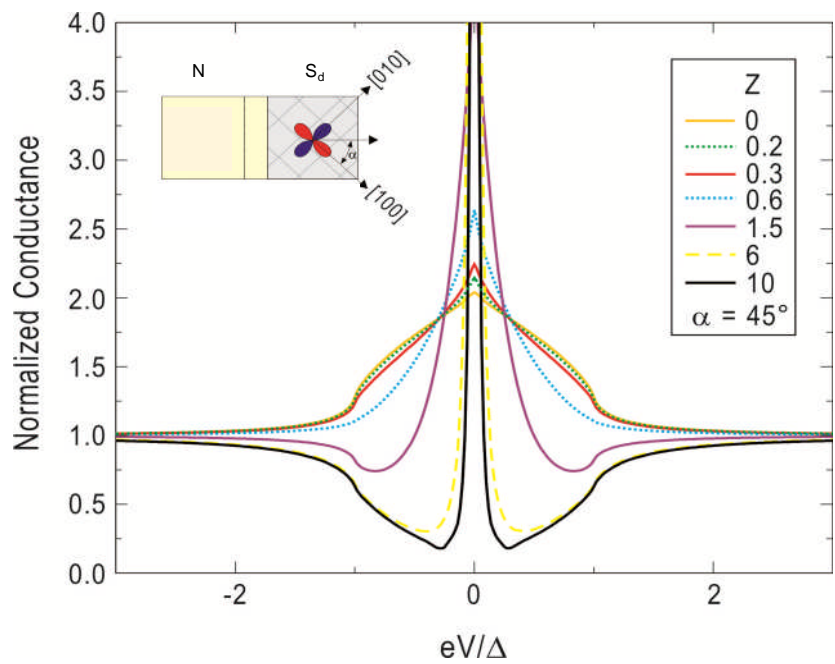

Fig. 4 (b)

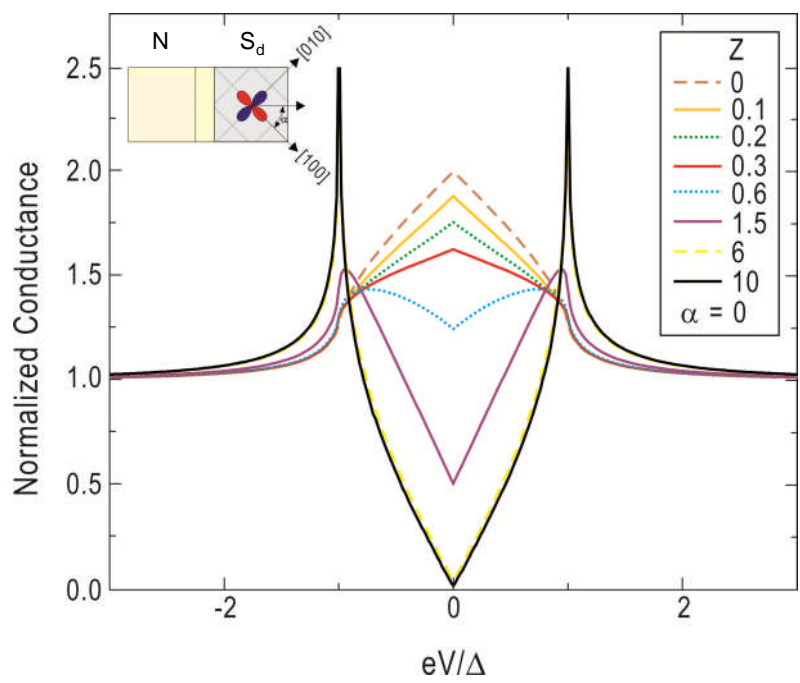

---

Figure 4  Calculated results of normalized differential conductance $(dI/dV)_S/(dI/dV)_N = (G_S/G_N)$ vs the ratio of bias voltage to gap $eV/\Delta$ as a function of barrier height Z based on a BTK model generalized to the case of normal metal / d-wave superconductor planar N/Sd tunnel junctions [26], where $S_d$ is a $d_{x2-y2}$ superconductor electrode). As indicated in the inset, the tunneling direction is defined as the angle, α, between the principal in-plane crystallographic axis and the norm of the interface boundary line. The tunneling is in the [110] direction with α = 45 deg, which is equivalent to the [100] direction of a $d_{xy}$ superconductor. The conductance spectra in Fig.4 (b) are for the cases of tunneling along [100] i.e. α = 0.

---

It is interesting to note that in a narrow range of 0 ≤ Z < 0.2, the conductance curves all show a dome-like feature centered at zero bias in the regions of -1 < $eV/\Delta$ < +1 with the maximum height of the conductance is roughly about twice of the conductance quantum $2G_0$ as expected that from the Andreev reflection and the normal reflection processes each contributes $G_0 = 2e^2/h$ to the conduction across the normal / superconductor (d or s wave). As the value of Z increases beyond 0.2, the dome-like feature gradually disappears and is replaced by the coherence peaks formed at $eV/\Delta$ = +1 and -1 for the case of α = 0; or the dome-like feature can be evolved into a sharp ZBCP with a height at zero bias much larger than 2 for the case of α = 45 deg, signaling the fact that the $N/S_d$ Josephson junctions are in the tunneling regime.  As it will be seen in the following discussion, it is exactly that these unique features of the BTK models make it possible to decipher information about the MBS component in such a complicated experimental conductance spectrum.

The theoretical predictions of simplest generalization of BTK for junctions consisting of d-wave superconductors [26] shown in Fig. 4(a) and (b) can be used as a guide for analyzing the experimental results presented in Fig. 3 (a), (b) and (c). Let's consider the temperature evolution of the conductance spectra of the $N/S_d$ tunnel junctions.  As a function of decreasing temperature, a dome-like feature in Fig. 3 (b) and 3 (c) emerges slightly below $T_c$. This allows for a somewhat accurate estimate of the value of the temperature-independent barrier height parameter Z. We can reasonably expect the low-Z dome-like features are also not sensitive to temperature variations. For the time being, we will focus only on the relative heights of the ZBCP in these conductance spectra and ignore the sub-gap oscillations near liquid helium temperatures. According to the theory [Kashiyawa and Tanakain, and others in ref. 26] as presented in Fig. 4 (a) and (b), a concave upward temperature dependent conductance suggests Z< 0.2 which in turn means the normalized differential conductance ($G_S/G_N$, the conductance ratio of the superconducting and normal states) at the top of the dome should be around 2, the maximum amount of charge transport of the combined normal and Andreev reflections across the N/S interface. To help in understanding the intriguing features of the low-temperature conductance, data in Fig. 3 (c) and Fig. 3 (b) are plotted in reduced units of $G_S/G_N$ and $eV/\Delta$ and presented in Figure 5 (a) and (b) respectively.

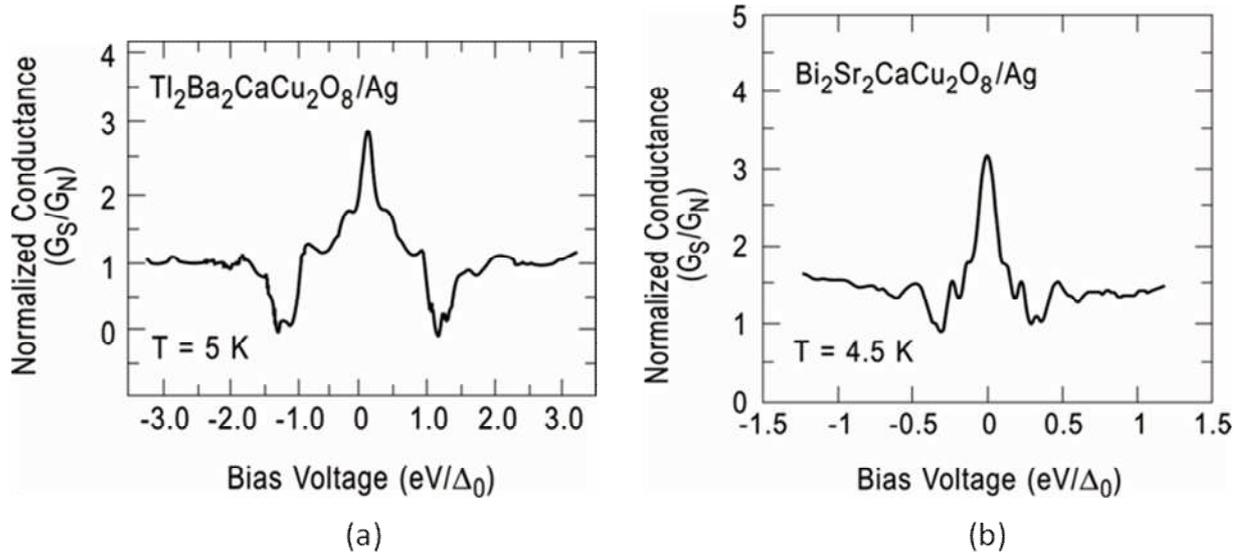

---

Figure 5 Bias voltage ($eV/\Delta_0$) dependence of the normalized conductance ($G_S/G_N$) of two d-wave superconductor/normal metal junctions at liquid helium temperatures: The original data of the junctions $Tl_2Ba_2CaCu_2O_8$ (c-axis oriented epitaxial film)/Ag in (a), and $Bi_2Sr_2CaCu_2O_8$(single crystal)/Ag in (b) were taken from references [33] and [32] respectively. Both of these two sets of results met the two criteria discussed in the text: 1) high-quality junction interface, and 2) in the low Z limit of the BTK model. Nominally the tunneling direction in (a) is [110]. In actuality, it is highly sensitive to the interface roughness on the nanometer scale because a short in-plane coherence length ( $\xi \approx$ 2 nm ) is characteristic of all cuprate superconductors. The normalized bias voltage $eV/\Delta_0$ is only a relative energy scale with $\Delta_0$ = 32 meV for $Bi_2Sr_2CaCu_2O_8$, and 20 meV for $Tl_2Ba_2CaCu_2O_8$. The fact that conductance dip in Fig. 5 (a) occurs at $eV/\Delta_0 \approx$ 0.5 instead of $\approx$ 1 as in Fig. 5 (b) may reflect that more relevant are the energies of the surface states at the interface [13].

---

There is a remarkable resemblance between the two conductance curves in Figure 5. The spectral similarity between the (001) $Tl_2Ba_2CaCu_2O_8$/Ag (Fig. 5a) and the (110) $Bi_2Sr_2CaCu_2O_8$/Ag (Fig. 5b) data is consistent with that the junctions in Fig. 3(c) and Fig. 5 (a) are probably dominated by in-plane Andreev reflection. As noted above, this domination follows directly from the two-dimensional nature of cuprates, which inherently favors in-plane over c-axis junction transmission when both types of junction facets are present, as may be expected on the rough surface of the as-grown (001) $Tl_2Ba_2CaCu_2O_8$ films. In addition, an intriguing feature in the both conductance spectra is that there are pronounced dips in conductance near the rise and fall of the dome-like broad peak. It is important to note that, as shown in Fig. 4(a) and (b), a BTK model for various $N/S_d$ tunnel junctions never calls for such a dip (right next to the central dome) below the normal state baseline in conductance spectrum. However, such unusual feature of conductance dip at the two sides of a domelike feature was indeed predicted for an in-plane planar $N/S_p$ junction[see Fig. 3(a) in Kashwaya et al. of [40], and Fig. 2 in Wu and Saokhin [41]. Here N is a normal metal such as Au or Ag, and $S_p$ stands for a junction electrode made of a two-dimensional chiral p-wave superconductor film. The theory also suggests that the conductance dip at a normalized bias voltage $eV/\Delta_0 = 1$ is more pronounced with an increasing strength of SOC at the junction (see Fig. 4 In ref. 41). Since the normal junction electrodes of the $N/S_d$ junctions are both made of Ag films, one can speculate that silver electrode may have induced a very strong SOC at the $Ag/S_d$ interface. This is possible in view of the well-established experimental observation of giant SOC effects with thin films of Au or Ag at various metallic surfaces and interfaces [C. R. Ast et al. (2007) in ref. 22]. The combined observation of a dome like feature **and** a sharp drop from the conductance base line is a clear evidence for the Rashba SOC induced mixing of $d_{xy}$ and $p_x + ip_y$ pairing states as predicted by theory [21]. This is a rather significant finding because it demonstrates, for the first time, a novel and simple electron transport measuring technique can be used to establish experimentally the two dimensional Rashba SOC effect at a junction interface.

Another remarkable characteristic shared by the conductance spectra in Fig. 5 (a) and (b) is that the total conductance at zero bias voltage can be approximately described by the following expression:

$$G_S/G_N = [G_{BTK} + G_{MF}] / G_N = 3 \qquad at\ V = 0 \qquad (2),$$

where $G_S/G_N$ is the conductance in the superconducting state normalized by that of the normal state, $G_{BTK}$ is the sum of the conductance through all charge transport channels as predicted by a generalized BTK for $d_{xy}$ and $p_x + i\ p_y$ pairing states[26, 40, 41], and $G_{MF}$ is conductance via the MBS channel at the junction interface [13, 21, 42, 6]. The fact that a dome-like feature was observed in the temperature evolution of the conductance spectra in both Fig. 3 (b) and (c), has to put the estimated Z value in the near zero regime(i.e. $Z \leq 0.3$) for these junctions. Thus, $G_{BTK}$ should be about $2\times 2e^2/h$ multiplied by a multi-channel factor which depends on the junction configuration and the micro-structural details of the junction interface. It is probably reasonable to assume that this factor does not change significantly as a function of temperature at low temperatures. Therefore it is expected that the ratio of $G_{BTK}/G_N$ is $\approx 2$ leading to the obvious conclusion that $G_{MF}/G_N \approx 1$ as predicted from the theory [13, 21, 42, 6] for quantized signature for Majorana fermions.

In short, the observation of angular resonant Andreev reflection **and** quantized ZBCP suggests the existence of MBS at the interface of the Ag (Au) / d-wave superconductor planar/ramp-type ballistic ($Z \approx 0$) junctions with tunneling in the [110] (or equivalently [100] in the $d_{xy}$ case) direction of the superconducting electrode. This is a direct consequence of Rashba SOC induced mixing of $d_{xy}$ and $p_x+ip_y$ pairing symmetries [13, 21, 42] as we emphasized before. Moreover, it has been well established in several theoretical studies [9, 10, 42-44] that, in the weak-coupling regime, a spinless p-wave pair state is generically characterized by MBS at zero energy in quasiparticle excitation spectrum. Such a Majorana zero mode is also bound to a half quantum vortex[9, 10, 42].

which is characterized by a time reversed doubly degenerate ground states. It is the ground state degeneracy of the MBS-hosting half quantum vortices that leads to non-Abelian statistics[9,10].

Therefore, for demonstrating non-Abelian statistics and eventualy building a prototype topological quantum computer, one needs to construct a superconducting heterostructure for hosting, manipulating and reading out the Majorana bound states. In another words, one needs a platform that is equipped with both HQV and MBS and, in addition, with adiabatic controls for varying and measuring the quantum states in the HQV- MBS ensemble.

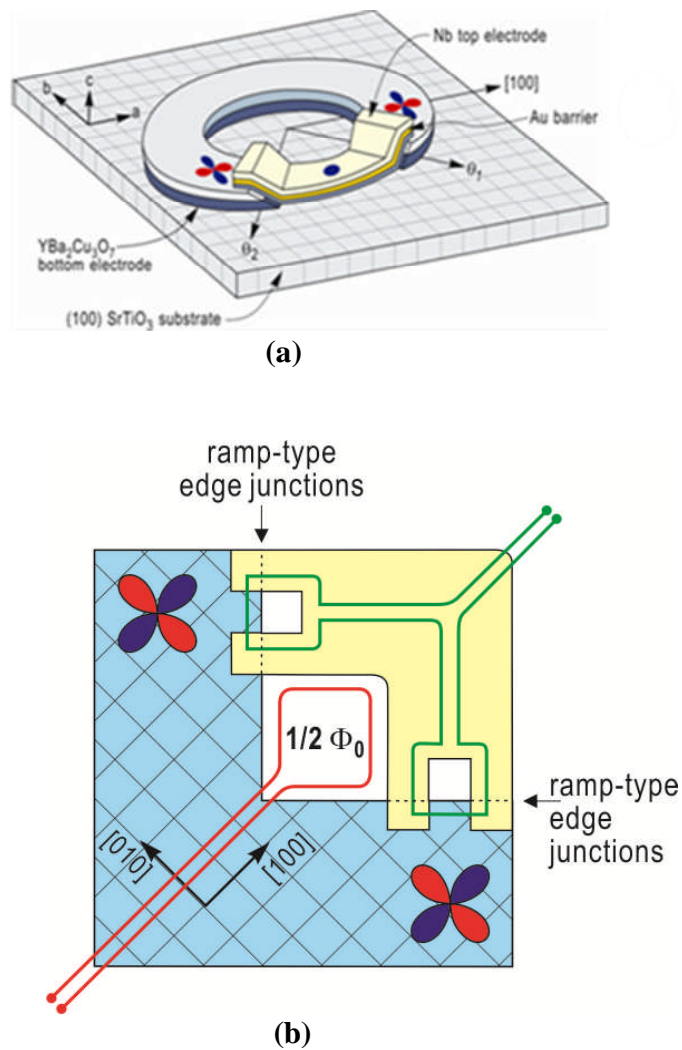

(a)

(b)

---

Fig. 6  (a) A schematic of the hetrosructure consists of two ramp-type edge junctions, depicted in Fig 1 (b), for example Nb/Au/YBCO on a (100) SrTiO3 substrate with the corresponding layer thickness 150 nm/20 nm/350 nm respectively.  The ramp edge angles $\theta_1$, $\theta_2$ are measured with respect to the crystallographic direction [100] of the c-axis oriented YBCO film. Except at the junction tunneling area, the upper surface of the bottom electrode is covered with 70 nm thick $SrTiO_3$ layer. The value of $\theta_1$ can be set at $-\pi/4$ (i.e. the longitudinal ramp edge is perpendicular to the [110] nodal direction), $\theta_2$ should be varied to maximize the resonant Andreev reflection as shown in Figure 2. If the junction interface is perfectly smooth, the optimized $\theta_2$ should be $-3\pi/4$. The inside (outside) diameter is 30 (120) µm. In Fig. 6 (b): a schematic of a proposed Majorana platform based on a d-wave π-loop consisted of two tunable Josephson ramp-type edge junctions, and two inductive flux control coils. The red coil is for controlling the eigenstates and parity of the qubit, and the green is for adjusting the tunnel barrier of the two tunable Josephson junctions.

---------------------------------------------------------------------------------------------------------------

With these operational requirements in mind, we propose in the following as a Majorana platform [see Fig. 6 (a) and (b)] based on a d-wave superconductor-based π-loop configuration that is made of two $S/NS_d$ ramp junctions like the ones shown in Fig. 1 (b). A π-loop is a superconducting ring-like structure interrupted by deliberately-oriented multiple Josephson junctions to achieve an odd number of sign changes in the supercurrent circulating in the loop. An example of π-loops is shown in Fig. 6 (a). As a result, it has a built-in intrinsic phase shift of π and is characterized by a doubly-degenerate time-reversed half flux quantum state (± ½ $\Phi_0$) in the ground state. It was used in phase-sensitive experiments to establish the d-wave pairing symmetry of cuprate superconductors [25] and in a variety of superconducting devices and circuits [20]. A superconducting loop with a π-phase shift in the orbital pair wavefunction always gives rise a HQV, in dependent of d-wave or p-wave pairing symmetry. However, in line with the discussion at the beginning of this paper, the zero-energy modes constructed from a superposition of hole-like and electron-like qp states in a spinless p-wave superconductor is a MBS and is topologically protected. Although the same construct without a Rashba spin-orbit coupling [21] superimposed at the junction interface in a pure d-wave system is topologically trivial.

To measure and manipulate the magnetic flux states and the Majorana zero modes attached to the HQV threading the two-junction ring, one needs to equip a standard π-loop, depicted in Fig. 6(a), with two tunable Josephson junctions, each consisted of two ramp edge junctions of the type shown in Fig.1 (b), and in addition two inductively coupled flux control coils. The green coil is used to control the magnitude of the critical supercurrent $I_c$, hence tunnel barrier of a double-well potential, generally used for describing the flux quantum states of such systems [45,46,25]. The red coil is for controlling the amount of magnetic flux threading through the center loop. In recent years, such superconducting circuit based qubits for non-topological quantum computation have been well-developed in theory and experiment [45, 46]. To use the setup depicted in Fig. 6(b) as a building block for constructing a viable Majorana qubit for topological quantum computing requires a careful consideration of issues related to adiabatic braiding operations of qubits, architecture of quantum logic gates and their readout schemes, - - . All these topics are crucial to the success of such Majorana-platform based computer and would be left for a future study.

# Acknowledgements

The author wishes to gratefully thank the following colleagues for useful comments and suggestions: S. Bravyi, A. Brinkman, C. –T. Chen, C.C. Chi, D. P. DiVincenzo, L. Fu, W. J. Gallagher, J. M. Gambetta, H. Hilgenkamp, M. B. Ketchen, D. –H. Lee, P. A. Lee, C. B. Lirakis , Y. Liu, S. S. Mahajan, C. –Y. Mou,  J.  Mannhart , D. M. Newns, J. Z. Sun, M Steffen, Graeme Smith, J.A. Smolin, J.Y.T. Wei, and S. K. Yip. Special thanks go to C. Schuppler for providing a copy of his PhD thesis.